\documentclass[12pt]{emsprocart}

\contact[dothanh@math.tamu.edu]{Ngoc T. Do, Mathematics Department, Texas A\& M University, 77843, College Station, Texas, USA}

\contact[kuchment@math.tamu.edu]{Peter Kuchment, Mathematics Department, Texas A\& M University, 77843, College Station, Texas, USA}

\contact[Beng.Ong@cgg.com]{Beng Ong, CGG, Houston, Texas, USA}
\usepackage{cite,amssymb,color}

\usepackage{graphicx}

\newtheorem{theorem}{Theorem}[section]
\newtheorem{lemma}[theorem]{Lemma}

\theoremstyle{definition}
\newtheorem{definition}[theorem]{Definition}

\theoremstyle{theorem}

\theoremstyle{corollary}
\newtheorem{corollary}[theorem]{Corollary}

\theoremstyle{conjecture}
\newtheorem{conjecture}[theorem]{Conjecture}

\theoremstyle{proposition}
\newtheorem{proposition}[theorem]{Proposition}

\theoremstyle{remark}
\newtheorem{remark}[theorem]{Remark}

\numberwithin{equation}{section}

\newcommand{\N}{\mathbb{N}}

\newcommand{\Z}{\mathbb{Z}}
\newcommand{\C}{\mathbb{C}}

\newcommand{\G}{\Gamma}

\newcommand{\bd}{\begin{definition}}
\newcommand{\ed}{\end{definition}}
\newcommand{\bt}{\begin{theorem}}
\newcommand{\et}{\end{theorem}}
\newcommand{\bl}{\begin{lemma}}
\newcommand{\el}{\end{lemma}}
\newcommand{\bc}{\begin{corollary}}
\newcommand{\ec}{\end{corollary}}
\newcommand{\bcon}{\begin{conjecture}}
\newcommand{\econ}{\end{conjecture}}
\newcommand{\br}{\begin{remark}}
\newcommand{\er}{\end{remark}}
\newcommand{\bp}{\begin{proposition}}
\newcommand{\ep}{\end{proposition}}
\newcommand{\be}{\begin{equation}}
\newcommand{\ee}{\end{equation}}
\newcommand{\bef}{\begin{figure}}
\newcommand{\eef}{\end{figure}}
\newcommand{\bea}{\begin{eqarray}}
\newcommand{\eea}{\end{eqarray}}
\newcommand{\ba}{\begin{array}}
\newcommand{\ea}{\end{array}}

\newcount\theTime
\newcount\theHour
\newcount\theMinute
\newcount\theMinuteTens
\newcount\theScratch
\theTime=\number\time
\theHour=\theTime
\divide\theHour by 60
\theScratch=\theHour
\multiply\theScratch by 60
\theMinute=\theTime
\advance\theMinute by -\theScratch
\theMinuteTens=\theMinute
\divide\theMinuteTens by 10
\theScratch=\theMinuteTens
\multiply\theScratch by 10
\advance\theMinute by -\theScratch

\def\today{{\number\day\space
 \ifcase\month\or
  January\or February\or March\or April\or May\or June\or
  July\or August\or September\or October\or November\or December\fi
 \space\number\year}}

\title[Resonant spectral gaps]{On resonant spectral gaps in quantum graphs\\ \emph{Dedicated to the 70th birthday of Pavel Exner}}

\author[N. Do, P. Kuchment, B. Ong]{Ngoc T. Do, Peter Kuchment, Beng Ong \thanks{The work of first two authors was partially supported by the NSF DMS grant \# 1517938.}}
\begin{document}



\begin{abstract}
In this brief paper we present some results on creating and manipulating spectral gaps for a (regular) quantum graph by inserting appropriate internal structures into its vertices. Complete proofs and extensions of the results are planned for another publication.
\end{abstract}



\maketitle

\section*{Introduction}

\begin{figure}[ht!]
\begin{center}
\scalebox{.7}{\includegraphics{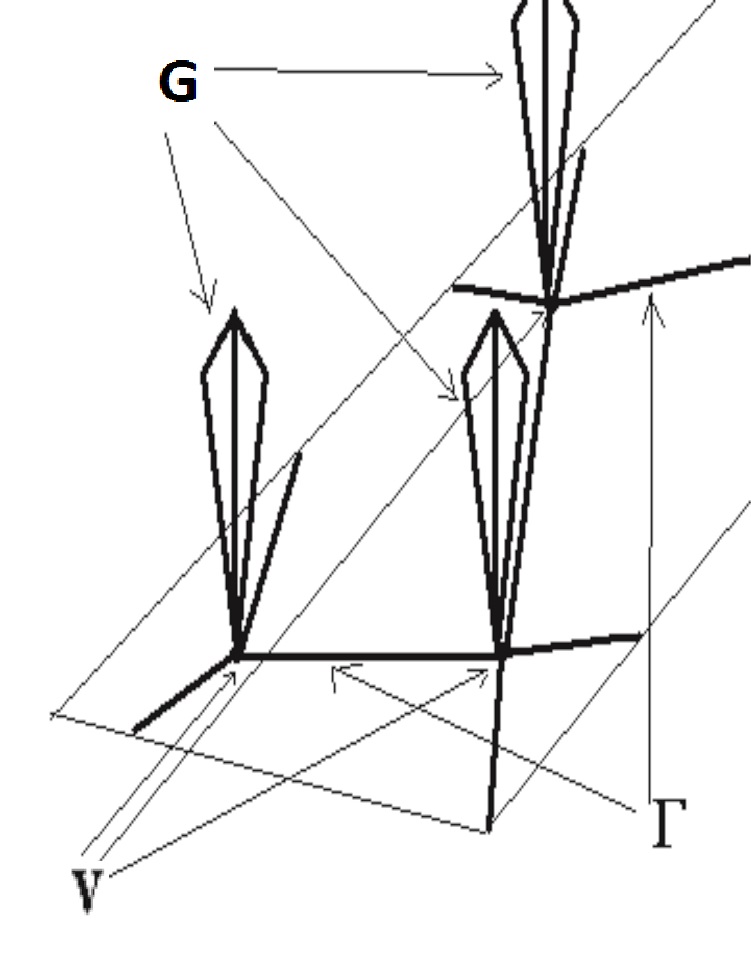}}
\caption{Decorations used by Schenker and Aizenman in \cite{SchAiz_lmp00}.}\label{F:decor}
\end{center}
\end{figure}

\begin{figure}[ht!]
\begin{center}
\scalebox{1.5}{\includegraphics{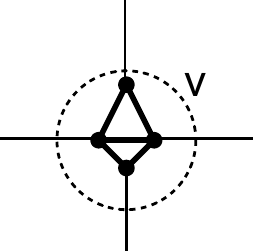}}
\caption{A ``spider'' decoration replacing a vertex $V$.}\label{F:spider}
\end{center}
\end{figure}

Existence of gaps in the spectra of operators of mathematical physics plays important role in many areas (e.g., solid state physics \cite{AshcroftMermin_solid}, photonic crystal manufacturing \cite{Joannopoulos_photocrystals,Kuch_pbg}, and in expander graphs construction \cite{WidgZig_anm02}, see also discussion in \cite[Section 6.1]{Kuch_BAMS}).
This also applies to constructions of thin branching structures (e.g., quantum wire circuits), which can be modeled by the so called quantum graphs (see \cite{BerKuc_book,AGA}). One of the standard ways to achieve the band-gap structure of the spectrum is by making the medium periodic, where the gaps may arise due to the Bragg scattering \cite{AshcroftMermin_solid}. However, existence of spectral gaps in periodic media is not guaranteed and is not easy to achieve and manipulate (see, e.g., \cite[Section 6.1]{Kuch_BAMS}). Thus, a different, resonant gaps technique has been explored, where identical resonators are distributed throughout the medium to create spectral gaps (the earliest reference known to the authors is \cite{Pavlov}). This idea was implemented in the discrete situation in \cite{SchAiz_lmp00} by attaching to each vertex $v$ of the graph $\G$ (which is the medium in this case) an identical \emph{decoration} (resonator) $G$ (Fig. \ref{F:decor}).
This technique has been extended to the case of quantum graphs, see \cite[Section 5.1 and references therein]{BerKuc_book}. However, it would be more convenient in many instances (e.g., in photonic crystal theory when considering the so called inverse opal structures) to insert some internal structure into each vertex, rather than attach a decoration (resonator) to it sideways. In other words, one is looking for a \emph{spider decoration} (Fig. \ref{F:spider})\footnote{Compare with the first step of the zig-zag construction of an expander \cite{WidgZig_anm02}.}.

Here one hits a snag. The nice procedure in \cite{SchAiz_lmp00} does not work nearly that well when the common ``boundary'' between $\G$ and $G$ consists of more than one point, as it is the case in the spider decorations.

When the boundary is a single point, by applying a rather standard technique used in considering the transmission problem between two media, one can rewrite the spectral problem on the decorated graph as the one on the original graph $\G$ with an additional energy (spectral parameter) dependent potential (\emph{Dirichlet-to-Neumann operator} of the decoration, see \cite[Section 5.1]{BerKuc_book} for details). Poles of this potential arise at the spectrum of the decoration, which leads to the gap opening.

With the boundary consisting of more than one point, the arising potential term is now a meromorphic matrix function, whose poles may or may not show up, depending on the vector the matrix function is applied to. Thus, it is easy to construct examples when a spider decoration does not lead to spectral gap opening \cite[Chapter 3]{Ong_diss}.

However, as it is shown in \cite[Chapter 3]{Ong_diss}, there are some special decoration constructions that resolve this problem in the case of finite graphs. The goal of this paper is to extend the (unpublished) considerations of \cite[Chapter 3]{Ong_diss}.

We provide necessary notions and notations in Section \ref{S:prel}, the auxiliary study of the Dirichlet-to-Neumann operator of a graph in Section \ref{S:aux}, the main result on gap opening  in Section \ref{S:main}, and conclusions and some remarks in Section \ref{S:remarks}.

\section{Preliminaries}\label{S:prel}

We consider a metric graph $\G_0$, i.e. a graph such that each its edge $e$ is equipped with a finite positive ``length'' $l_e$ and a coordinate $x$ identifying it with the segment $[0,l_e]$ (see more detailed discussion in \cite[Chapter 1]{BerKuc_book}). We will use the standard notations $V(\G_0)$ and $E(\G_0)$ for the sets of vertices and edges of the graph respectively.

We will assume in this text that the following condition on the edge lengths of the graphs $\G_0$ is satisfied:
 \be\label{E:uniform}
 \mbox{there exists }l>0\mbox{ such that } l\leq l_e\leq 1/l<\infty, \forall e\in E(\G_0).
  \ee
 In particular, this is true when the graph $\G_0$ is either finite (i.e., has a finitely many edges and vertices) or periodic (i.e. is equipped with the co-finite action of the group $\Z^p$ for some $p>0$, see \cite[Section 4.1]{BerKuc_book}).

Let us also assume that $\G_0$ is a $d$-regular graph (i.e. the degree of each of its vertices is equal to $d$) and $G$ is a finite metric graph with at least $d$ vertices and a singled out subset $B\subset V(G)$ consisting of $d$ vertices. The set $B$ will be called the \emph{boundary} of $G$. For each vertex $v\subset V(\G_0)$ we establish a 1-to-1 correspondence between the edges adjacent to $v$ and the elements of $B$. One can now decorate in a natural way the vertex $v$ with the internal structure, which is a copy of $G$ (see again Fig. \ref{F:spider}). Doing this for all vertices of $\G_0$, we obtain the \emph{decorated graph} $\G$.

All graphs we consider are equipped with the self-adjoint operators\footnote{Usually called Kirchhoff Laplacians.}, $H_0$ in $L^2(\G_0)$ and $H$ in $L^2(\G)$, as follows: on each edge they act as $-d^2/dx^2$, with the domain consisting of functions $f$ such that
\begin{enumerate}
\item $f\in H^2(e)$ for each edge $e$;
\item $f$ is continuous on the whole graph;
\item at each vertex, the sum of the outgoing derivatives of $f$ along all adjacent edges is equal to zero (Kirchhoff condition);
\item the sum $\sum\limits_{e}\|f\|^2_{H^2(e)}$ is finite (automatic for a finite graph).
\end{enumerate}
Here $H^2(e)$ is the standard Sobolev space on the segment $e=[0,l_e]$.

We also denote by $H_G$ the analogous operator on $G$, with the exception that at the boundary vertices $v\in B$, Dirichlet conditions $f(v)=0$ are imposed instead of Kirchhoff ones. The spectrum $\sigma({H_G})$ of this operator is discrete \cite[Theorem 3.1.1]{BerKuc_book}.

We denote by $\Sigma_D$ the (discrete under our conditions of finiteness or periodicity of the graph) set of Dirichlet eigenvalues of all edges of $\G_0$, i.e.
\be
\Sigma_D:=\{\left(n\pi l_e^{-1}\right)^2\}_{n\in\N, e\in E(\G_0)}.
\ee

Let us also denote by
\be
N: \bigoplus\limits_{e\in E(G)}H^2(e) \rightarrow l^2(B)
\ee
the \emph{Neumann} operator that for any function
$$
f\in \bigoplus\limits_{e\in E(G)}H^2(e)
$$
and a vertex $v\in B$ produces the value at $v$ equal to the sum of the outgoing derivatives of $f$ along the edges of $G$ adjacent to $v$. Here we denote by $l^2(B)$ the $d$-dimensional Hilbert space of functions on $B$.

We can now define, for any $\lambda\notin \sigma(H_G)$, the \emph{Dirichlet-to-Neumann operator} (in fact, a $d\times d$-matrix) $\Lambda(\lambda)$ as follows: for any $\phi\in l^2(B)$ let $u$ be the (existing and unique) solution of the following problem:
\be\label{E:inhom}
\begin{cases}
-d^2u/dx^2=\lambda u \mbox{ on each } e\in E(G),\\
u \mbox{ satisfies continuity and Kirchhoff condition at each vertex }v\notin B,\\
u|_B=\phi.
\end{cases}
\ee
Then,
\be
\Lambda(\lambda)\phi:=Nu.
\ee
It is clear that $\Lambda(\lambda)$ is a meromorphic function with poles at $\sigma(H_D)$ only (see \cite[Section 3.5]{BerKuc_book} for more detailed consideration of Dirichlet-to-Neumann operator in the quantum graph case and its relation to the resolvent of $H_G$).

\section{Auxiliary statements}\label{S:aux}

Let $\lambda_0\in\sigma(H_G)$. As it was indicated before, and as we will see clearly in the next section, it will be important for us that for any non-zero $\phi\in l^2(B)$ the vector function $\Lambda(\lambda)\phi$ still has a pole at $\lambda_0$. It is clearly sufficient to consider vectors $\phi$ that belong to the unit sphere $S$ of $l^2(G)\approx \C^d$.

The following auxiliary result is crucial for our goal:

\bt\label{T:aux}\cite{Ong_diss}
\indent
\begin{enumerate}
\item If for a given $\phi\in S$ and $\lambda=\lambda_0$ the problem \ref{E:inhom} has a solution, then $\Lambda(\lambda)\phi$ does not have singularity at $\lambda_0$.
\item If the problem \ref{E:inhom} has no solution for $\lambda=\lambda_0$ and any $\phi\in S$, then for any $\phi$, the following estimate holds in an (independent of $\phi$) neighborhood of $\lambda_0$:
\be\label{E:estim}
\|\Lambda(\lambda)\phi\| \geq \frac{C}{|\lambda-\lambda_0|}\|\phi\|
\ee
with a constant  $C$ independent of $\phi$.
\end{enumerate}
\et
Thus, we will be looking at graphs $G$ with boundary $B$ such that the problem \ref{E:inhom} has solution only for zero Dirichlet data $\phi$. Rather than trying to describe all graphs that have this property, we will provide (for any size $d$ of the boundary $B$) constructions when this does happen, which will be sufficient for our purpose of gap opening.
\bt\label{T:suffic}
Let $l_0>0$ and $n$ be an odd natural number. Suppose that the pair $G,B$ satisfies the following conditions:
\begin{enumerate}
\item Graph $G$ contains a cycle\footnote{which can be assumed non-self-intersecting.} $Z$ consisting of an odd number of edges of the length $l_0$;
\item Each boundary vertex $v\in B$ either belongs to $Z$, or is connected to a vertex of $Z$ by a path of edges of length $l_0$ each.
\end{enumerate}
Then, for $\lambda_0=(n\pi/l_0)^2$, there exist a neighborhood $U$ of $\lambda_0$ and a constant $C$ such that the inequality (\ref{E:estim}) holds for any $\phi$ and $\lambda\in U$.
\et
\begin{proof}
WLOG, let us assume that $Z$ is non-self-intersecting and consider an edge $e\in Z$ (of length $l_0$, as all edges in the cycle). The solution of (\ref{E:inhom}) for $\lambda=\lambda_0$ on this edge has the form $$
u=a\cos\left(\dfrac{n\pi}{l_0}x\right)+b \sin\left(\dfrac{n\pi}{l_0}x\right).
$$
At the endpoints $x=0, l_0$, the second term vanishes. Since $n$ is odd, the first term changes its value from $a$ to $-a$ at these points. Going around an odd cycle, one concludes that this is possible only if $a=0$, and thus $u$ vanishes at all vertices of $Z$. In particular, it vanishes at all boundary vertices that belong to $Z$. For $v\in B\setminus Z$, as there exists a path of edges of length $l_0$ from $v$ to a vertex from $Z$, where we know that $u$ vanishes, the same consideration shows that $u$ vanishes at all vertices of the path. Therefore, the problem (\ref{E:inhom}) does not have solution for non-zero $\phi$, and thus, due to Theorem \ref{T:aux}, the inequality (\ref{E:estim}) follows.

\end{proof}

\section{The main result}\label{S:main}

We are ready now to formulate and sketch the proof of the main result of this article:
\bt\label{T:main}
Let $l_0>0$ and $n$ be an odd natural number. Let also the $d$-regular graph $\G_0$ satisfy (\ref{E:uniform}), and finite graph $G$ with boundary $B$, $|B|=d$ and the decorated graph $\G$ are defined as before. Suppose that the following conditions are satisfied:
\begin{enumerate}
\item $\lambda_0=(n\pi/l_0)^2\notin \Sigma_D$, with $dist(\lambda_0,\Sigma_D)=r>0$;
\item The decoration (resonator) $G, B$ satisfies the conditions of Theorem \ref{T:suffic}.
\end{enumerate}
Then there exists a punctured neighborhood of $\lambda_0$, depending on $G$, topology of $\G_0$, and $r$ only, which does not belong to the spectrum $\sigma(H)$.
\et
\begin{proof}

To understand the idea of the proof, let us assume first that the graph $\G_0$ (and thus $\G$) is finite. Then the spectrum of $H$ is discrete. Thus, if $\lambda\in\sigma(H)$, there exists a non-zero eigenfunction $u$. Assume that the neighborhood of $\lambda_0$ has radius less than $r_1<r$. Then it contains no elements of $\Sigma_D$.

Removing all internal edges and vertices from each of the decorations in $\G$, one gets a disjoint union of edges of $\G_0$, since each former vertex $v\in \G_0$ is replaced by $d$ vertices $v_1, \dots, v_d$, see Fig \ref{F:dots}.
\begin{figure}[ht!]
\begin{center}
\scalebox{.7}{\includegraphics{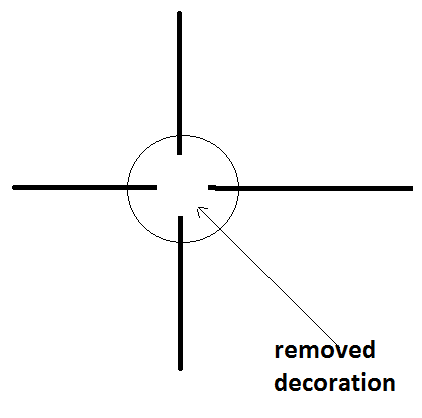}}
\caption{Decoration removed from a former vertex of degree $4$.}\label{F:dots}
\end{center}
\end{figure}

We denote this new graph $\widetilde{\G}$.

Denoting
$$
\phi_v:=(u(v_1),\dots,u(v_d))^t,\,\phi^\prime_v:=(du/dx_{e_1}(v_1),\dots,du/dx_{e_d}(v_d))^t,
$$
where $e_1,\dots,e_d$ are the edges formerly adjacent to vertex $v$ and coordinates $x_{e_j}$ increase from the value zero at vertex $v$, one can rewrite the equation for $u$ on $\G$ as the following one on $\widetilde{\G}$:
\be\label{E:removed}
\begin{cases}
-u^{\prime\prime}
=\lambda u \mbox{ on each edge},\\
\phi^\prime_v=-\Lambda(\lambda)\phi_v \mbox{ at each vertex } v\in V(\G_0).
\end{cases}
\ee
Since $\lambda_0$ is at a qualified distance from the Dirichlet spectrum $\Sigma_D$, the resolvent estimates for self-adjoint operators together with embedding theorems show that
\be\label{E:resolvest1}
\sum_e\|\phi\|^2_{H^2(e)}\leq M \sum_v\|\phi_v\|^2,
\ee
\be\label{E:resolvest2}
\sum_v\|\phi^\prime_v\|^2\leq M \sum_v\|\phi_v\|^2
\ee
for some constant $M$ depending only on the distance from the Dirichlet spectrum and topology of $\G_0$. On the other hand, if $\lambda$ is sufficiently close to $\lambda_0$ (how close, depends on $M$ and the decoration $G$), according to Theorem \ref{T:aux}, one has
\be\label{E:oppositeest}
\sum\limits_v\|\phi^\prime_v\|^2\geq \frac{C^2}{|\lambda-\lambda_0|^2}\sum\limits_v\|\phi_v\|^2> M \sum\limits_v\|\phi_v\|^2.
\ee
This contradiction proves the claim of the theorem for finite graphs.

In the case of infinite graphs, generalized eigenfunctions with control of growth need to be used \cite{Lenz_gener_eig,HislPost}. The details will be provided elsewhere, but so far we illustrate this on the simplest example of periodic graphs,

In the case of a periodic graph, assumption that $\lambda$ belongs to the spectrum of $H$ implies existence of a quasi-periodic Bloch-Floquet generalized eigenfunction $\phi(x)$, which under the $\Z^p$-shifts acquires only a phase shift (see details in \cite[Section 4.2]{BerKuc_book}). Then the above consideration for the finite graph goes smoothly, if the summation over edges and vertices of $\G_0$ is replaced by the same for the compact orbit space graph $\G_0/\Z^p$.

Another option in the periodic case is to use the Floquet-Bloch decomposition and apply the above (finite case) argument for each value of the quasi-momentum\footnote{See \cite[Chapter 4]{BerKuc_book} for these notions and constructions.}.
\end{proof}

\section{Conclusions and final remarks}\label{S:remarks}
\begin{enumerate}
\item Although it is probably not easy to understand the case of a general ``spider'' decoration, the main result of this article allows one to create spectral gaps rather easily at prescribed locations. Indeed, the value $(n\pi l_0^{-1})^2$ involves an arbitrary positive length $l_0$ and odd natural number $n$, which gives one a significant freedom of choosing location. As soon as this is done, one can easily produce a spider decoration $G$ that achieves the goal. For instance, if $d=4$, each degree $4$ vertex can be replaced with the structure shown in Fig. \ref{F:repl}, where we created an odd cycle through three boundary vertices and connected the fourth one to them with a single edge.
\begin{figure}[ht!]
\begin{center}
\scalebox{0.5}{\includegraphics{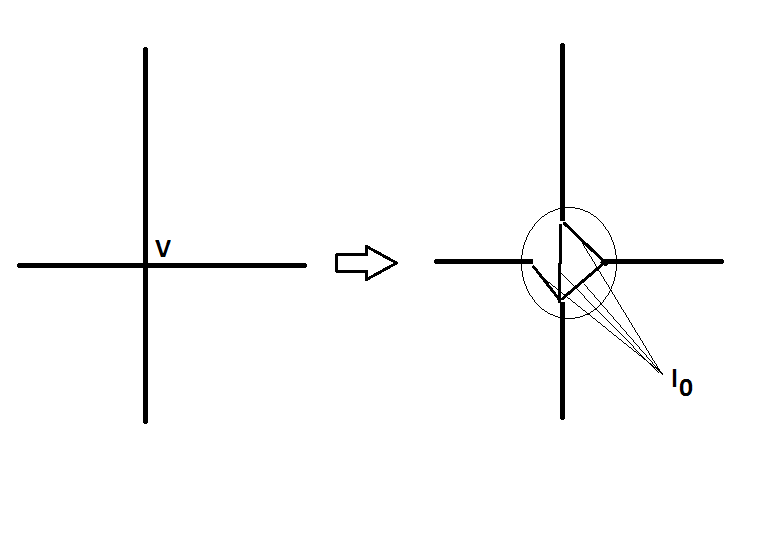}}
\caption{A degree $4$ vertex replaced by a ``spider'' with edges of lengths $l_0$.}\label{F:repl}
\end{center}
\end{figure}
\item It was shown in \cite{Ong_diss} on examples that for even cycles and/or even $n$ the claim of the Theorem \ref{T:main} is incorrect.
\item The regularity condition on the graph $\G_0$ is not truly necessary, at least in the case of finite and periodic graphs. Indeed, one can manage variable degrees by choosing different decorations, adjusted to each particular vertex and such that the corresponding resonant values $\lambda_0$ agree.
\item Besides gap's existence at a given location, its size is of importance. It depends on the value of the constant $C$ in (\ref{E:estim}). Our construction allows for a variety of decorations achieving the gap at the same location. It would be interesting to analyse the dependence of $C$ on the decoration, to pick the most effective designs.
\item More general vertex conditions can be considered and Kirchhoff conditions are used just for simplicity.
\end{enumerate}

\section*{Acknowledgment}
The work of the first two authors was partially supported by the NSF DMS grant \# 1517938. The authors express their gratitude to the NSF for the support. They are also grateful to the referees for useful comments.

\bibliography{MKperiodic}{}
\bibliographystyle{amsplain}

\end{document}